\documentstyle[12pt]{article}

\def \we {\wedge}

\newcommand{\rf}[1]{(\ref{#1})}

\renewcommand{\thefootnote}{\fnsymbol{footnote}}

\newcommand{\newsection}{    
\setcounter{equation}{0}
\section}
\def\appendix#1{
  \addtocounter{section}{1}
  \setcounter{equation}{0}
  \renewcommand{\thesection}{\Alph{section}}
  \section*{Appendix \thesection\protect\indent \parbox[t]{11.715cm} {#1} }
  \addcontentsline{toc}{section}{Appendix \thesection\ \ \ #1}
  }

\def \N {{\cal N}}

\def \foot {\footnote}
\def \bi{\bibitem}
\def \la {\label}

\def \CC{{\cal C}}
\def \ov {\over}

\def \I {{\cal E}}
\def \J {{\cal J}}

\def \L{{\widehat{L}}}

\def\nline{\,\nabla\kern -0.7em\raise0.2ex\hbox{/}\,\,}
\def\yline{\,y\kern -0.47em /}
\def\aline{\,a\kern -0.49em /}
\def\parline{\,\partial\kern -0.55em /\,\,}

\def\apr{{a^\prime}}
\def\bpr{{b^\prime}}
\def\cpr{{c^\prime}}
\def\dpr{{d^\prime}}
\def\alpr{{\alpha^\prime}}
\def\bepr{{\beta^\prime}}
\def\gapr{{\gamma^\prime}}
\def\depr{{\delta^\prime}}

\def\aha{{\hat{a}}}
\def\bha{{\hat{b}}}
\def\cha{{\hat{c}}}

\def \t {\theta}
\def \s{\sigma}

\def\NPB#1(#2)#3{{\it Nucl. Phys.} {\bf B#1} (#2) #3}
\def\PRD#1(#2)#3{{\it Phys. Rev.} {\bf D#1} (#2) #3}
\def\PLB#1(#2)#3{{\it Phys. Lett.} {\bf B#1} (#2) #3}

\def\RMP#1(#2)#3{{\it Rev. Mod. Phys.} {\bf #1} (#2) #3}
\def\MPLA#1(#2)#3{{\it Mod. Phys. Lett.} {\bf A#1} (#2) #3}
\def\CQG#1(#2)#3{{\it Class. Quantum Grav.} {\bf #1} (#2) #3}
\def\AP#1(#2)#3{{\it Ann. Phys.} {\bf #1} (#2) #3}
\def\SJNP#1(#2)#3{{\it Sov. J. Nucl. Phys.} {\bf #1} (#2) #3}

\def\np {{Nucl. Phys. }}
\def \pl {{Phys. Lett. }}

\def \del{\partial}

\def \FF {{\rm F}}

\def\det{\hbox{det}}
\def\be{\begin{equation}}
\def\ee{\end{equation}}

\def \ci {\cite}

\def \ka {\kappa}

\def \ads {$AdS_5\times S^5\ $}
\begin{document}

\begin{titlepage}
\begin{flushright}
FIAN/TD/98-23   \\
Imperial/TP/97-98/53  \\
hep-th/9806095\\
\end{flushright}
\vspace{.5cm}

\begin{center}
{\LARGE  Supersymmetric  D3 brane   action  in
  AdS$_5 \times $S$^5$  }\\[.2cm]
\vspace{1.1cm}
{\large R.R. Metsaev${}^{{\rm a,b}}$\footnote{\ E-mail: metsaev@lpi.ac.ru}
and A.A. Tseytlin${}^{{\rm a,b}}$\footnote{\ E-mail: tseytlin@ic.ac.uk} }\\
\vspace{18pt}
${}^{{\rm a\ }}${\it
Department of Theoretical Physics, P.N. Lebedev Physical
Institute,\\ Leninsky prospect 53, 117924, Moscow, Russia
}\\
${}^{{\rm b\ }}${\it Blackett Laboratory,
Imperial College, London SW7 2BZ, U.K.}
\end{center}
\vskip 0.6 cm

\begin{abstract}
We  find  the  space-time supersymmetric
and  $\kappa$-invariant   action for a D3-brane propagating in the  
  $AdS_5\times S^5$ 
background.  As in the previous  construction of
 the fundamental string action in this  maximally supersymmetric
string vacuum the    starting point 
is   the  corresponding superalgebra $su(2,2|4)$. 
We  comment on  the super Yang-Mills interpretation of the
gauge-fixed form of the  action.
\end{abstract}

\end{titlepage}
\setcounter{page}{1}
\renewcommand{\thefootnote}{\arabic{footnote}}
\setcounter{footnote}{0}

\newsection{Introduction}
The action of a D3-brane  probe propagating in a curved 
type IIB supergravity background is described  by a 
combination of a Born-Infeld-type term and a Wess-Zumino type term
(see, e.g., \ci{polch,leigh,douglas,schmid,tset}). 
The  supersymmetric and $\kappa$-invariant expressions for the  D3-brane action 
in flat space and general type IIB backgrounds were constructed in 
\ci{ced,aps,bt}. 

Type IIB supergravity has two maximally supersymmetric 
vacua: flat space and \ads background \ci{S2}. 
In the case of the flat space the D3-brane  action
 computed in
the  static gauge and $\kappa$-symmetry gauge is a non-linear 
generalisation of the abelian  $\N=4$ supersymmetric  Yang-Mills 
action which has 16 linearly  and 16 non-linearly realised 
supersymmetries \ci{aps}. 
Since  \ads is the large charge or near-horizon      limit  \ci{gibb}
of the D3-brane solution of
\ci{stro,duff}, the corresponding  
action  may be interpreted as describing  a D3-brane 
probe propagating near the core of a D3-brane 
source. 

 In  string theory,  a collection of $N$ parallel 
D3-branes is described  by  connecting  open strings \ci{polch},  or
 (at low energies) by $U(N)$ SYM theory \ci{wii}.
The action for a single D3-brane  separated from $N-1$ others  
is obtained by  integrating out massive 
 open strings stretched between the `probe' and the `source' 
(see, e.g., \ci{dougl}). In   the large $N$ (large charge) limit the BI-type D3-brane  action 
should  essentially coincide  with 
the leading IR  part  
of  the {\it quantum}  $\N=4$  SYM effective action 
obtained by keeping  the $U(1)$ $\N=4$  vector multiplet as an external 
 background and 
integrating out  massive SYM  fields
 \ci{chep,maldstr,malda,keski}. 
Since the quantum $\N=4$, $D=4$  SYM theory is conformally 
invariant, the resulting action 
should  also have (spontaneously broken by scalar field background and thus 
non-linearly realised) conformal symmetry.
The non-linear conformal invariance  of the  bosonic part of the 
static-gauge D3-brane 
action  in \ads was demonstrated  in \ci{malda,kal1,ckkt}.

Our aim will be to find the  complete supersymmetric
(invariant under the  32 global supersymmetries of
the \ads vacuum) 
 and $\kappa$-invariant 
 form of 
the D3-brane action in \ads space.
After fixing the static gauge and $\kappa$-symmetry gauge
the action  will   have a  `SYM effective action' interpretation.
Like the flat space action, it  will have 16 linear  and 
16 non-linear (conformal)  supersymmetries.
Its conformal  invariance is a  consequence of the $SO(4,2)\times SO(6)$
isometry of the  \ads metric
and 
 is manifest {\it before} the  static 
gauge fixing.

The knowledge of such supersymmetric 
action is  quite non-trivial  
in view of its non-polynomiality and 
the absence of the manifestly supersymmetric
 $\N=4$ (superfield)  formalism:
one effectively determines  the  exact 
supersymmetry transformation laws
to all orders in low-energy expansion. 
The fermionic structure of the action  is  of interest
also  
 in connection with  recent discussions 
of the SYM--supergravity correspondence 
(see, e.g., \ci{kleb,gkt,malda,gkp,wit,datr,ferr,deal}).

One possible approach to the  construction of the  action 
is to start with  the general type IIB background actions  in 
\ci{ced,aps,bt} and plug in the  type IIB superfields
representing the  \ads  vacuum 
(the corresponding supergeometry was recently  discussed in \ci{kalkum,krr}). 
This approach, however, is  indirect and  somewhat  complicated 
as it   does not  make explicit use of   the basic 
symmetries of the problem.

Our strategy  will be  instead  to use directly  the supergroup
$SU(2,2|4)$ which is the underlying symmetry 
of the \ads string vacuum.\footnote{Since the $U(1)$ symmetry 
relating the two supercharges of IIB supergravity is {\it not} 
a symmetry of type IIB string theory \ci{gree,howe},
one is to omit  this $U(1)$ factor
  from the $U(2,2|4)$
symmetry \ci{gunay}  of the supergravity vacuum. 
The corresponding superalgebra 
we  use in this  paper does not contain the $U(1)$ and central charge generators.
In mathematics literature this 
superalgebra   is denoted 
by $psu(2,2|4)$)  \ci{kaplan}, but we will use 
its common `simplified' notation.}
As in the  previous construction of the type IIB string action in 
\ads  \ci{mt},  we  shall  obtain 
the  space-time  supersymmetric  and 
$\kappa$-symmetric  D3-brane action in terms of the 
invariant  Cartan
one-forms defined on 
the coset superspace $SU(2,2|4)/[SO(4,1)\otimes SO(5)]$.
Our  method is conceptually very close to the one 
used in \cite{aps} to find 
the action of a  D3-brane propagating  in flat space 
as  a $D=4$  `Born-Infeld plus Wess-Zumino' -type model on the
flat 
 coset superspace
($D=10$ super Poincare group)/($D=10$ Lorentz group).

As in \cite{mt}, our starting point   will be  the superalgebra
$su(2,2|4)$
containing the  two pairs of
 translations and rotations, 
 $(P_a, J_{ab})$ for $AdS_5$ and $(P_\apr,J_{\apr\bpr})$
 for $S^5$,  and the  supersymmetry generators (32 odd translations)
 which  form  the two
 $D=10$ Majorana-Weyl spinors
 $Q^{\alpha \alpr}_{ 1,2}$ of the same chirality. 
Our notation  and conventions will  be close 
to those of  \ci{mt} and are explained in Appendix A
where we also  write down  the commutation relations for $su(2,2|4)$. 
Appendix B contains some basic relations for Cartan forms $L^A$
 on 
the coset superspace.

\newsection{Cartan forms  }
To find the manifestly  super-invariant and
 $\kappa$-invariant  D3-brane action we will use  the formalism of
Cartan forms  on the coset superspace $SU(2,2|4)/[SO(4,1)\otimes SO(5)]$.
The left-invariant Cartan  1-forms
$$L^A = dX^M L^A_{M}\ ,  \qquad \ \ \
X^M=(x, \theta)\,,$$
are defined by
\begin{equation}\label{expan1}
G^{-1}dG 
=L^\aha P_\aha+\frac{1}{2}L^{\aha \bha}J_{\aha\bha}
+L^{\hat{\alpha}}Q_{\hat{\alpha}} \ ,
\end{equation}
where $G= G({x,\theta})$ is a coset representative in $SU(2,2|4)$.
$L^a$ and $L^\apr$ are the 5-beins,  $L^{\hat{\alpha}1,2}$ are the two
Majorana-Weyl spinors and  $L^{ab}$ and $L^{\apr\bpr}$ ($L^{a\apr}=0$)
are the Cartan connections (for their detailed form see \ci{mt}
and Appendix B). They satisfy 
the  Maurer-Cartan equations  (B.1)--(B.4) implied by the
structure of the  $su(2,2|4)$ superalgebra.

 A specific choice of
$G(x,\theta)$ which we shall use in this paper is 
\be
G=g(x)e^{\theta Q}\,,\qquad
\ee
where $g(x)$ is a coset representative of
$[SO(4,2)\otimes SO(6)]/[SO(4,1)\otimes SO(5)]$, i.e.
  $x=(x^\mu,x^{\mu'})$ provides a 
certain parametrization of $AdS_5\times S^5$ which may be kept 
arbitrary.\foot{The  use of  a
 concrete parametrization for $\theta$ is needed, however, 
to find  the representation for the 2-form 
${\FF}$  which enters the BI action (see below). 
As in the flat space case  \ci{aps}, 
${\FF}$ cannot be expressed in terms of the  Cartan forms  only.}
Then 
\be
G^{-1}dG=e^{-\theta Q}De^{\theta Q}\,,
\ee
where $D$ is the closed  covariant differential  
\be
D=d+\frac{1}{2}\omega^{\aha\bha}J_{\aha\bha}
+e^\aha P_\aha\,,\ \ \ 
\qquad 
D^2=0\,. 
\ee
The explicit form of $D\theta$ is 
 \be
D\theta=(d+\frac{1}{4}\omega^{\aha\bha}\Gamma^{\aha\bha}
-\frac{{\rm i}}{2}\I e^\aha\sigma_+\Gamma^\aha)\theta\,,
\ee
where the matrices $\sigma_+$ and $\I$ are defined in Appendix A.

Let us note that in the 32-component spinor 
notation used in the present paper the superstring 
action in \ads background  found in  \ci{mt}
has the following  form  valid in an arbitrary parametrisation ($\del M_3=M_2$)
\be
I=-\frac{1}{2}\int_{M_2}  d^2\sigma\  \sqrt{g}\ g^{ij}
\  L_i^\aha L_j^\aha 
+\int_{M_3}  H_3 \ , \ \ \ \ 
\la{acti1}
\ee
$$
\ \ 
H_3 = {\rm i}
L^\aha\we\bar{L}\Gamma^\aha\we {\cal K}L \ . $$
In the explicit parametrisation (2.2)
\be
I=-\frac{1}{2}\int_{M_2}  d^2\sigma\  \bigg[ \sqrt{g}\ g^{ij}
\  L_i^\aha L_j^\aha  + 
4{\rm i}\epsilon^{ij}  \int_0^1 dt\  L_{it}^\aha\ 
\bar{\theta}{\Gamma}^\aha {\cal K} L_{jt} \bigg]\ ,
\la{acti2}
\ee
where 
the notation are explained in \rf{noo},\rf{noot}.

\newsection{D3-brane action  }

The D3-brane action  depends on the coset 
superspace coordinates $X^M=(x^{\hat m},\theta)$ and vector field
strength $\del_i A_j -\del_j A_i$. As in \ci{ced,aps,bt}, 
it 
is given by the  sum of the BI and WZ  terms
\begin{equation}\label{action}
S=S_{\rm BI} +S_{\rm WZ}\,,
\end{equation}
where (we set the 3-brane tension to  be 1 and $M_4=\del M_5$)
\be
S_{\rm BI}=-\int_{M_4}  d^4\sigma\ \sqrt{-\det(G_{ij}+{\FF}_{ij})}\ , 
\la{hhh}
\ee
\be 
S_{\rm WZ} =\int_{M_4} \Omega_4 = \int_{M_5} {H}_5 \ , 
\ \ \ \ \ \ \ \  \ \ \  {H}_5=d \Omega_4 \ .    
\la{hdo}  
\ee
The  induced world-volume metric $G_{ij}$ is 
($i,j=0,1,2,3$)
\be
G_{ij}=L_i^\aha L_i^\aha = \del_i X^M  L_M^\aha \del_j X^N L_N^\aha
\,, \ \ \ \ \ \ \ \ \  \ \  
L^\aha (X(\s)) = d\sigma^i L_i^\aha\,.
\ee
 The supersymmetric extension 
${\FF}=\frac{1}{2}{\FF}_{ij}d\sigma^i\wedge d\sigma^j$  
   of the 
world-volume gauge field
strength  2-form 
$dA$  is  found to be  
\begin{equation}\label{fexp}
{\FF}=dA +2{\rm i}\int_0^1 dt\  L_t^\aha\we  
\bar{\theta}\Gamma^\aha {\cal K} L_t \  ,  
\end{equation}
where 
$L_t^\aha(x,\theta)\equiv L^\aha(x,t\theta), 
\ \ 
L_t(x,\theta)\equiv L(x,t\theta).
$
The $\theta$-dependent correction term  in \rf{fexp}
given by the integral over the auxiliary parameter $t$
is exactly  the same 2-form as in the string action 
\rf{acti2}.
This representation corresponds to the 
specific choice of coset representative made above. 
Note that while ${\FF}$ is not expressible in terms 
of Cartan forms only, 
 its exterior derivative 
is
\be
d{\FF}={\rm i}\bar{L}\we \L \we{\cal K} L \,, \ \ \ \ \ \ \ \ \ \ 
\L\equiv L^\aha \Gamma^\aha\,.
\la{dee}
\ee
This important 
 formula can be proved my making use of the 
Maurer-Cartan equations and 
equations (\ref{p1})--(\ref{p3}) from Appendix B. 

As a result, $d{\FF}$ is manifestly invariant under supersymmetry, 
and then  so is ${\FF}$, 
provided  one defines   appropriately   the  
transformation
of $A$ to cancel the  exact variation of the second (string WZ) term 
in   \rf{fexp}  (cf. \ci{ced,aps}).  Note that this is
 the same transformation 
that is needed  to make  the  superstring action \rf{acti2}
defined on a {\it disc}  and coupled to $A$ at the boundary
invariant under supersymmetry.

As in flat space \ci{ced,aps}, the  super-invariance 
of $S_{\rm WZ}$ follows  from supersymmetry of 
the closed 5-form  $H_5=d\Omega_4$.
We shall determine the supersymmetric
$H_5$ from the requirement
 of $\kappa$-symmetry of the full action $S$  which    fixes  this 5-form 
uniquely.  The $\kappa$-transformations 
are defined by  (see \rf{fff}) 
\be
\delta_\kappa x^\aha=0\,,\ \ 
\qquad \delta_\kappa \theta =\kappa\,,
\ee
where the  transformation parameter satisfies the constraint
\begin{equation}\label{kappa}
\Gamma \kappa =\kappa\,, \qquad \ \ 
\Gamma^2=1\,.
\end{equation}
  $\Gamma$  is  given   by 
\be
\Gamma=\frac{\epsilon^{i_1\ldots i_4}}{\sqrt{-\det(G_{ij}+{\FF}_{ij})}}
\bigg(\frac{1}{4!} \Gamma_{i_1\ldots i_4} \I
+\frac{1}{4}\Gamma_{i_1i_2}{\FF}_{i_3i_4} \J
+\frac{1}{8}{\FF}_{i_1i_2}{\FF}_{i_3i_4} \I\bigg)
\,, \ee
where
\be
\Gamma_{i_1\ldots i_n}\equiv \L_{[i_1}\ldots \L_{i_n]}\,,\ \ \ \ \  \ \ \ 
\qquad 
\L_i\equiv L_i^\aha \Gamma^\aha
\,.\ee
The corresponding variation of the metric $G_{ij}$ is
\begin{equation}\label{kvg}
\delta_\kappa G_{ij}=-2{\rm i}\delta_\kappa\bar{\theta}(\L_i L_j+\L_j L_i)\ .
\end{equation}
The variation of ${\FF}$  is given by
\be
\delta_\kappa {\FF}
=2{\rm i}\delta_\kappa \bar{\theta} \L\we {\cal K} L
\,, \ee
or,  in components, 
\begin{equation}\label{kvf}
\delta_\kappa
{\FF}_{ij}=2{\rm i}\delta_\kappa \bar{\theta}(\L_i {\cal K} L_j-\L_j {\cal K} L_i)\,.
\end{equation}
Our main  statement is that the D3-brane 
action $S$  in (3.1) is $\kappa$-invariant provided
the   5-form ${H}_5$ is given by
$$
{H}_5
={\rm i}\bar{L}\we \bigg(\frac{1}{6}\L\we\L\we\L \I+{\FF}\we \L \J           \bigg)   \we  L
$$
\begin{equation}\label{h5}
+\ \frac{1}{30}\bigg(\epsilon^{a_1\ldots a_5}L^{a_1}\we \ldots\we L^{a_5}
+ \epsilon^{\apr_1\ldots \apr_5}
L^{\apr_1}\we \ldots\we  L^{\apr_5}\bigg)
\,.\end{equation}
It is possible to check (using Maurer-Cartan equations and Fierz identities) 
 that ${H}_5$ is closed, i.e. the equation \rf{hdo} 
 is consistent
and thus determines  $S_{\rm WZ}$.

The important fact is that  ${H}_5$ is expressed  in terms 
of the 
Cartan 1-forms 
and super-invariant  ${\FF}$ only. 
This implies that ${H}_5$ is invariant under space-time
supersymmetry.
Then from (\ref{hdo}) we conclude that $\delta_{susy} \Omega_4$
is exact, so  that the WZ term  
\rf{hdo},  like the BI  term \rf{hhh},  is supersymmetry invariant.

To put the fermionic part of the WZ term in the action in a more explicit 
form let us make a rescaling $\theta \to t\theta$ and define
\be
H_{5t} \equiv H_5|_{\theta \rightarrow t \theta}\ , \ \ \ \ \ \ \ \ 
{\rm F}_t\equiv {\rm F}|_{\theta\rightarrow t\theta}
\, .  \ee
Since  
$ L(x,tt'\theta)=
L_t(x,t'\theta)=L_{tt'}(x,\theta)
$
one can show that (cf. \rf{fexp},\rf{dee}) 
\be
{\rm F}_t=dA+2{\rm i}\int_0^t dt' \, \bar{\theta}\widehat{L}_{t'} \wedge {\cal K } L_{t'} \ , \ \ \ \ \ \ \ 
\partial_t {\rm F}_t=2{\rm i}\bar{\theta}\widehat{L}_t\wedge {\cal K}L_t \ . 
\la{iim}
\ee
Then using the  defining equations for the Cartan forms \rf{p1}--\rf{p3} 
one finds from \rf{h5} the following differential equation
\be 
\partial_t H_{5t}
=d\bigg[ 2{\rm i}(\frac{1}{6} \bar{\theta}\widehat{L}_t\wedge\widehat{L}_t\wedge \widehat{L}_t {\cal E} L_t
+\bar{\theta}\widehat{L}_t\wedge {\rm F}_t\wedge{\cal J}L_t)\bigg]
\,,\ee
which determines the $\theta$-dependence of $H_5$. 
With the initial condition 
\be
(H_{5t})_{t=0} = H_5|_{\theta=0} = H_5^{(bose)} =
\frac{1}{30}(\epsilon^{a_1\ldots a_5} e^{a_1}\wedge \ldots \wedge e^{a_5}
+ \epsilon^{\apr_1\ldots \apr_5} 
e^{\apr_1}\wedge \ldots \wedge e^{\apr_5}) \ 
, 
\ee
where $e^a$ and $e^{a'}$ are (pull-backs of) the  vielbein forms  of 
$AdS_5$ and $S^5$,    
   we conclude  that  $H_5=(H_{5t})_{t=1}$  is given by  $d\Omega_4$ (see \rf{hdo}) 
  where 
\be
\la{exx}
\Omega_4 = 2{\rm i} \int_0^1 dt \,
\bigg(\frac{1}{6} \bar{\theta}\widehat{L}_t\wedge \widehat{L}_t\wedge \widehat{L}_t {\cal E} L_t
+\bar{\theta}\widehat{L}_t\wedge{\rm F}_t\wedge {\cal J}L_t\bigg)
+ \Omega_4^{(bose)} \,. \ee
The explicit form of the $\theta$-independent part
$\Omega_4^{(bose)}$  (satisfying $d\Omega_4^{(bose)}=H_5^{(bose)}$) 
depends on a particular choice of coordinates on \ads.
Thus  the WZ term in \rf{hdo} can be written as  (cf. \rf{acti2})
\be
S_{\rm WZ} =  2{\rm i} \int_{M_4}  \int_0^1 dt
 \,
\bigg(\frac{1}{6} \bar{\theta}\widehat{L}_t\wedge \widehat{L}_t\wedge \widehat{L}_t {\cal E} L_t
+\bar{\theta}\widehat{L}_t\wedge{\rm F}_t\wedge {\cal J}L_t\bigg)
+ \  \int_{M_5} H_5^{(bose)} \ . 
\la{fina}\ee
Using \rf{con1},\rf{con2} one can then find the expansion of $S_{\rm WZ}$
in powers of $\theta$.

It is useful to  recall that the only non-trivial
background fields  in  \ads  vacuum  are
the space-time metric and the self-dual RR 5-form.
The bosonic parts of the last two terms in 
${H}_5$ \rf{h5}    represent, indeed, the 
standard bosonic couplings  of the D3-brane to the 5-form background
(their explicit coordinate form can be found, e.g., in \ci{ckkt}).
The action we have obtained  contains 
also the fermionic terms required to make this coupling supersymmetric and
$\kappa$-invariant. 

Let us stress again that we have  started with 
the BI action expressed in terms of the  Cartan 1-forms and the  2-form in
\rf{acti2},\rf{fexp} (determined in \ci{mt}) 
as implied by the structure of \ads space
 or the $su(2,2|4)$  superalgebra.  We then 
 fixed the form of ${H}_5$
from the requirement of $\kappa$-symmetry of the full action.
The fact that we have reproduced  the bosonic part of the 
self-dual 5-form 
 is in perfect agreement   with the result 
of \ci{ced,bt}  that  the 
D3-brane action is $\kappa$-invariant only in a 
 background   which is a  solution of IIB supergravity
(for  \ads space this  implies the presence of the  non-trivial 
self-dual 5-form field \ci{S2}).

\newsection{Remarks on  gauge fixing}
To summarize, we have
found  the supersymmetric  action for D3-brane probe
propagating in $AdS_5\times S^5$ background.
 The action is   given by (\ref{action})--(\ref{fexp}),\rf{fina}, 
with the  closed  5-form 
 defining the WZ term 
 given in (\ref{h5}).

This  action   is   world-volume reparametrisation invariant 
and $\ka$-invariant.  Its advantage is that it is manifestly 
invariant under the  symmetries of \ads vacuum:
bosonic isometries $SO(4,2) \times SO(6)$ and 32 supersymmetries.
However, it  does not have a particularly  simple form when written in 
terms of the 
coordinates $(x,\theta)$,  even 
using the 
closed expressions   for the 
 Cartan forms in terms of $\theta$ \ci{krr} given in 
(\ref{con1}),\rf{con2}.

 To put  the action  in a  
more explicit form  and  also 
to   establish a  connection to the SYM  theory  discussed in the Introduction
we need to 
 (i) choose  special  bosonic coordinates in \ads, \ 
(ii)  fix the  static gauge  so that the D3-probe is oriented parallel 
to the D3-source, 
  and  (iii) fix  the $\kappa$-symmetry gauge in a way that simplifies
the fermionic part of the action.
After fixing the  local symmetry gauges 
only the $ISO(3,1)\otimes SO(6)$  and 16 supersymmetries 
part of the original symmetry will
remain manifest, while the   superconformal symmetry
will be realised non-linearly.

One standard choice of the bosonic coordinates  in \ads is  such that 
$ds^2 = {x^2\ov R^2} dx_i dx_i  +  { R^2\ov x^2}  { dx_s dx_s }$,  where
$x^2 \equiv x_s x_s$, \ $s=1,...,6$,  and $R$ is the radius
(which is set equal to 1 in the rest of the paper). 
The static gauge choice is  then 
$x_i = \sigma_i$.\foot{The flat-space limit  (which is possible 
to take in the  D3-brane action {\it before} 
the static gauge choice) 
is obtained by  changing the coordinates  so that 
$x^2 = R^2 e^{- 2z/R}$, etc., and taking the limit $R\to \infty$.}

Both D3-brane actions -- in flat space  and in \ads space --
inherit the full set of the 32 supersymmetries 
of the corresponding type IIB supergravity vacua.  Their gauge-fixed forms, however, have
only   16 linearly realised supersymmetries. 
The interpretation of the remaining 16 supersymmetries 
as conformal  ones is possible only in the \ads case. 
Moreover, this interpretation seems to depend on 
a proper choice the $\ka$-symmetry gauge which 
should be different, e.g.,  from the $\theta_1=0$ choice 
in \ci{aps}. 
How to fix $\kappa$-symmetry gauge in the \ads action case 
in the most natural way is an important and 
 open  question. 
The  difference between the two actions
 is  related to the  fact that  while 
the flat space action has explicit scale $\alpha'$,
the role of such scale in the \ads action 
is played by the modulus of the scalar field.\foot{Let us note also 
that in contrast to the \ads one, 
the flat-space BI-type  D3-brane action
 is not related to quantum SYM theory -- the higher-order  terms in it
may be interpreted as  tree-level  string-theory $\alpha'$ corrections.}

A more complicated   open  problem  is a  non-abelian generalisation
of the  abelian D3-brane  action   we  have found. 
As the \ads action is different from  the  flat space one 
this problem  may  have a  different solution compared to 
the one  proposed  in \ci{TT}. 
 One obvious 
 suggestion -- to replace the fields 
$(x,\t,A)$   by $U(N)$ matrices and  to 
add the overall  symmetrised trace --
may not work as the  trace structure of the quantum SYM action 
 appears to be more involved (cf. \ci{chep}).

\setcounter{section}{0}
\setcounter{subsection}{0}
\begin{center}
{\bf Acknowledgments}
\end{center}
We are  grateful  to  
R. Kallosh
 for  useful  discussions. 
This  work was supported in part
by PPARC,   the European Commission  TMR programme  grant ERBFMRX-CT96-0045,
the INTAS grant No.96-538, 
 the Russian Foundation for Basic Research Grant No.96-01-01144
and the Royal Society visiting grant.

\appendix{Notation and conventions}
We use the following  conventions for
indices:
\begin{eqnarray*}
a,b,c=0,1,\ldots, 4 &\qquad &so(4,1) \hbox{ vector indices ($AdS_5$ tangent space)}\\
\apr,\bpr,\cpr=5,\ldots, 9 &\qquad & so(5) \hbox{ vector
indices ($S^5$ tangent space) }\\
\aha,\bha,\cha =0,1,\ldots, 9& & \hbox{ combination  of }
(a,\apr), (b,\bpr), (c,\cpr) \hbox{\ ($D=10$  vector indices)}
\\
\alpha,\beta,\gamma,\delta =1,\ldots,4 &\qquad &
so(4,1) \hbox{ spinor indices ($AdS_5$)}\\
\alpr,\bepr,\gapr,\depr =1,\ldots,4 &\qquad
&so(5) \hbox{ spinor indices ($S^5$) }\\
\hat \alpha,\hat \beta,\hat \gamma  =1,\ldots, 32& & 
\hbox{\ $D=10$  Majorana-Weyl spinor indices}
\end{eqnarray*}
The commutation relations of the  even part of $su(2,2|4)$ which is 
$so(4,2)\oplus so(6)$ are 
$$
[P_a,P_b]=J_{ab}\,,
\qquad\qquad\qquad
[P_\apr,P_\bpr]=-J_{\apr\bpr}\,,
$$
$$
[P_a,J_{bc}]=\eta_{ab}P_c-\eta_{ac}P_b\,,
\qquad
[P_\apr,J_{\bpr\cpr}]=\eta_{\apr\bpr}P_\cpr -\eta_{\apr\cpr}P_\bpr\,,
$$
$$
[J_{ab},J_{cd}]=\eta_{bc}J_{ad}+3\hbox{ terms}\,,
\qquad
[J_{\apr\bpr},J_{\cpr\dpr}]
=\eta_{\bpr\cpr}J_{\apr\dpr}+3\hbox{ terms}\,.
$$
In \cite{mt} the commutation relations for the  odd part were 
 written in terms  of 16-component spinor notation. It
turns out, however, that  calculations are 
 simplified if one uses the  32-component notation.
We shall use   the following representation for $32\times 32$ 
$\Gamma$-matrices 
\be
\Gamma^a=\gamma^a\otimes 1 \otimes \sigma_1\,,
\qquad
\Gamma^\apr = 1\otimes \gamma^\apr \otimes \sigma_2\,,
\qquad
\CC=C\otimes C^\prime \otimes {\rm i}\sigma_2
\,,\ee
where  $\CC$, $C$ and $C^\prime$ are the charge conjugation matrices for 
$so(9,1),$ $so(4,1)$ and $so(5)$ Clifford algebras respectively.
The 
Majorana condition  is 
$\bar \Psi = 
\Psi^\dagger \Gamma^0 =\Psi^T {\cal C}.
$
The 5d  Dirac matrices satisfy 
$
\gamma^{(a} \gamma^{b)}=\eta^{ab}=(-++++)\,,
\ \ 
\gamma^{(\apr} \gamma^{\bpr)}=\eta^{a'b'}=(+++++)
$ and 
$
\gamma^{a_1\ldots a_5}={\rm i}\epsilon^{a_1\ldots a_5}, \ 
$
$
\gamma^{\apr_1\ldots \apr_5}=\epsilon^{\apr_1\ldots \apr_5}. 
$

We shall 
 use the  following representation
for each of the two 32-component  Majorana-Weyl
(negative chirality)  supergenerators
 \be
Q^{\hat{\alpha}}=
\left(\begin{array}{c}
0\\
-Q^{\alpha\alpr}
\end{array}\right)\,, \ \ \ \ \ \ \ \  
Q_{\hat{\alpha}}\equiv Q^{\hat{\beta}} \CC_{\hat{\beta}\hat{\alpha}}
\,,
\ee
where $Q^{\alpha\hat{\alpha}}$ is a 16-component spinor. 
The two supergenerators   $(Q_{\hat{\alpha}}^1$,
$Q_{\hat{\alpha}}^2)$  of $su(2,2|4)$
will be combined into a 2-vector 
\begin{equation}\label{vec}
Q=\left(
\begin{array}{c}
Q^1\\
Q^2
\end{array}
\right) \,.
\end{equation}
The commutation relations for the odd part of the superalgebra  in \cite{mt} 
can be rewritten  as
$$
[Q,P_\aha]=\frac{{\rm i}}{2}Q\I \sigma_+ \Gamma_{\aha}\,,
$$
\be 
\{Q_{\hat{\alpha}},Q_{\hat{\beta}}\}
=-2{\rm i}(\CC\Gamma^\aha \pi_+)_{\hat{\alpha}\hat{\beta}}P_\aha
+\I\Bigl[(\CC\Gamma^{ab}\sigma_-)_{\hat{\alpha}\hat{\beta}}J_{ab}
-(\CC\Gamma^{\apr\bpr}\sigma_-)_{\hat{\alpha}\hat{\beta}}J_{\apr\bpr}\Bigr]
\,,\ee
where $\pi_+$ and $\sigma_\pm$ stand
 for $32\times 32$ matrices 
$1\times 1\times \pi_+$ and 
$1\times 1\times \sigma_\pm$, 
\be
\pi_+=\frac{1}{2}(1+\sigma_3)\,,\ \ 
\quad
\sigma_+=\frac{1}{2}(\sigma_1+{\rm i}\sigma_2)\,,\ \ 
\quad 
\sigma_-=\frac{1}{2}(\sigma_1-{\rm i}\sigma_2)\,,
\la{kkk}\ee
and  $\sigma_i$ are the usual Pauli matrices. 
 We shall also  use the following  $2\times 2$ matrices
\be
\I=\left(
\begin{array}{cc}
0  &  1\\
-1 &  0
\end{array}
\right)\,,
\qquad 
\J=\left(
\begin{array}{cc}
0  &  1\\
1   &  0
\end{array}
\right)\,,
\qquad
{\cal K}=\left(
\begin{array}{cc}
1   &  0\\
0   &   -1 
\end{array}
\right)
\,, \la{noo}\ee
which 
will  act on the two internal indices of $Q$  in (\ref{vec}).

The chirality of the spinor Cartan 1-forms 
 $L^{\hat{\alpha}1,2}$ and odd coordinates  $\theta^{\hat{\alpha}1,2}$ is 
opposite to that of  $Q^{\hat{\alpha}}$, i.e. in the  32-component notation 
$
L^{\hat{\alpha}}=\left(
\begin{array}{c}
L^{\alpha\alpr}\\
0
\end{array}
\right).
$
The $L^{1,2}$ and $\theta^{\hat{1,2}}$  can  be combined into 2-vectors
 like (\ref{vec}) and $L$ and $\theta$ in the text 
stand for such  vectors.

\appendix{Basic relations for Cartan forms on  coset superspace  }
The Cartan forms satisfy the  Maurer-Cartan equations 
implied by  the  $su(2,2|4)$ superalgebra
\be
dL^\aha=-L^{\aha\bha}\wedge L^\bha -{\rm i}\bar{L}\Gamma^\aha \wedge L
\,,\ee
\be
dL=\frac{{\rm i}}{2}\sigma_+ L^\aha \Gamma^\aha \wedge \I L
-\frac{1}{4} L^{\aha\bha} \Gamma^{\aha\bha}\wedge L
\,, \ \ \ \ \  
d\bar{L}=\frac{{\rm i}}{2}\bar{L} \I \wedge \Gamma^\aha  L^\aha \sigma_+
-\frac{1}{4} \bar{L}  \Gamma^{\aha\bha} \wedge L^{\aha\bha}
\,, \ee
\be
dL^{ab}=-L^a\wedge L^b-L^{ac}\wedge L^{cb}+\bar{L}\Gamma^{ab}\sigma_-\wedge \I L\,,
\ee
\be
dL^{\apr\bpr}=L^\apr \wedge L^\bpr -L^{\apr\cpr}\wedge L^{\cpr\bpr}
-\bar{L}\Gamma^{\apr\bpr}\sigma_-\wedge \I L\, . 
\ee
As in \ci{mt} we set the  radii   of $AdS_5$ and $S^5$   to be 1. 
It is often useful to use the following expressions for the  variations of 
Cartan forms which  are also implied by the
structure of the  $su(2,2|4)$
\be
\delta L^\aha=d\delta x^\aha +L^{\aha\bha}\delta x^\bha
+L^\bha\delta x^{\bha\aha}+2{\rm i}\bar{L}\Gamma^\aha \delta \theta\,,
\ee
\be
\delta L= d \delta \theta 
-\frac{{\rm i}}{2} \sigma_+ L^\aha \Gamma^\aha \I\delta \theta
+\frac{1}{4} L^{\aha\bha} \Gamma^{\aha\bha}\delta \theta
+\frac{{\rm i}}{2} \sigma_+ \delta x^\aha \Gamma^\aha \I L
-\frac{1}{4}\delta x^{\aha\bha} \Gamma^{\aha\bha} L
\,, \ee
\be
\delta \bar{L}= d \delta \bar{\theta} 
+\frac{{\rm i}}{2} \delta\bar{\theta} \I  \Gamma^\aha L^\aha\sigma_+ 
-\frac{1}{4} \delta \bar{\theta}\Gamma^{\aha\bha} L^{\aha\bha}
-\frac{{\rm i}}{2} \bar{L}\I \Gamma^\aha \delta x^\aha \sigma_+
+\frac{1}{4}\bar{L} \Gamma^{\aha\bha} \delta x^{\aha\bha} 
\,, \ee
where
\be
\delta x^\aha\equiv \delta X^M L_M^\aha\,,
\qquad
\delta x^{\aha\bha}\equiv \delta X^M L_M^{\aha\bha}\,.
\qquad
\delta \theta \equiv \delta X^M L_M^{\vphantom{5pt}}\,, \la{fff}
\ee
 Let us 
make  the rescaling
$\theta\rightarrow t\theta$ and introduce
\be
L_t^\aha(x,\theta)\equiv L^\aha(x,t\theta)\,, 
\qquad
L_t^{\aha\bha}(x,\theta)\equiv L^{\aha\bha}(x,t\theta)\,,
\qquad
L_t(x,\theta)\equiv L(x,t\theta)\,,
\la{noot}\ee
with the initial condition 
\be
L_{t=0}^\aha=e^\aha\,,
\qquad
L_{t=0}^{\aha\bha}=\omega^{\aha\bha}\,,
\qquad
L_{t=0}=0\,, 
\ee
where $e^\aha$, $\omega^{\aha\bha}$ are the 5-beins and the 
Lorentz connections
for $AdS_5\times S^5$. Then the defining equations 
 for the  Cartan 1-forms are (see eqs. (A.10)--(A.12) and (B.2)--(B.4)
in  \ci{mt} for details)
\begin{eqnarray}
\label{p1}
&&\partial_t L_t^\aha=-2{\rm i}\bar{\theta}\Gamma^\aha L_t\,,
\\
&&
\partial_t L_t=d\theta -\frac{{\rm i}}{2}\sigma_+\Gamma^\aha \I \theta L_t^\aha
+\frac{1}{4}\Gamma^{\aha\bha}\theta L_t^{\aha\bha}\,,
\\
\label{p3}
&&
\partial_t L_t^{ab}=2\bar{\theta}\I \Gamma^{ab}\sigma_- L_t\,,
\qquad
\partial_t L_t^{\apr\bpr}=-2\bar{\theta}\I \Gamma^{\apr\bpr}\sigma_-L_t\ .
\end{eqnarray}
One can find a closed  solution to these equations \ci{krr}
(we set   $t=1$) 
 \be
\label{con1}
L= V(\theta) D\theta\,, 
\ \ \ \ \ \ \ \ 
L^\aha=e^\aha-2{\rm i}\bar{\theta}\Gamma^\aha W (\theta) D\theta
\,, 
\ee
\be\la{con2}
L^{ab}=\omega^{ab}
+2\bar{\theta}{\cal E}\Gamma^{ab}\sigma_- W(\theta) D\theta
\,,
\ \ \ \ \  
L^{\apr\bpr}=\omega^{\apr\bpr}
-2\bar{\theta}{\cal E}\Gamma^{\apr\bpr}\sigma_- W(\theta) D\theta
\,, \ee
where the matrices  $V$ and $W$  are defined by 
\be
V\equiv \frac{\hbox{sinh}\sqrt{m}}{\sqrt {m}} = 1 + {1 \ov 3!}m +
{1 \ov 5!}m^2  +  ... \ ,
\ee
\be
W\equiv \frac{\hbox{cosh}\sqrt{ m} \ -1 }{m }
=  {1 \ov 2!} + {1 \ov 4!} m + {1 \ov 6!} m^2 +    ...  \ , 
\ee
and $m$ is   a  matrix  quadratic in $\theta$ 
\be
m=-\sigma_+\Gamma^\aha \I \theta\  \bar{\theta}\Gamma^\aha
+\frac{1}{2}\Gamma^{ab}\theta\ \bar{\theta}\I \Gamma^{ab}\sigma_-
-\frac{1}{2}\Gamma^{\apr\bpr}\theta\ \bar{\theta}\I \Gamma^{\apr\bpr}\sigma_-
\, . \ee
While the relations (B.1)--(B.7) are valid in  an arbitrary parametrisation
of the coset superspace,  (B.11)--(B.18) apply  only in the
 parametrisation of (2.2).

Note that in  many formal  calculations  it is more convenient
to use directly the defining equations (\ref{p1})--(\ref{p3}) rather
then the  solution
 (\ref{con1}),   the explicit expressions
\rf{con1},\rf{con2} may be useful in discussion of $\kappa$-symmetry gauge fixing
and  related  applications. 
As is well known, the   analogs of the expressions \rf{con1},\rf{con2}
can be written down for  the Cartan forms corresponding 
to a general symmetric space (see, e.g., \ci{brown}).


\end{document}